\documentclass{PoS}

\usepackage[utf8]{inputenc}
\usepackage{url}

\title{Tetraquark and the flux tube recombination}

\ShortTitle{Tetraquark and the flux tube recombination}

\author{\speaker{Marco Cardoso} \\%\thanks{A footnote may follow.}\\
        CFTP, Instituto Superior Técnico\\
        E-mail: \email{mjdcc@cftp.ist.utl.pt}}

\author{Pedro Bicudo\\
        CFTP, Instituto Superior Técnico\\
        E-mail: \email{bicudo@ist.utl.pt}}

\author{Nuno Cardoso\\
        CFTP, Instituto Superior Técnico\\
        E-mail: \email{nunocardoso@cftp.ist.utl.pt}}

\abstract{
Here we study the static potential for the two quarks and two antiquarks
system.
First this is done using the tetraquark operator, which has been previously used to calculate the static potential.
This is found to give good results in the region where the tetraquark is expected to be the ground state, however failing outside it.
To repair this, we resort to a variational method.
This let us study the first excited state besides the ground state of
the system in two different particle dispositions, one where the quarks
are on the same side of a rectangle and the other where they are at opposite sides.
Results for the field components and for the lagrangian density are presented.
}

\FullConference{International Workshop on QCD Green's Functions, Confinement and Phenomenology,\\
		September 05-09, 2011\\
		Trento Italy}

\begin{document}

\section{Motivation}

Systems constituted by two quarks and two antiquarks are of extreme importance for strong interaction physics.
Not only because they are a starting point for meson-meson scattering,
but also because of the possible existence of bound-states --- tetraquarks, initially predicted by Jaffe \cite{Jaffe:1976ig}.
There are several observed resonances who are candidates to tetraquarks.
The most recent candidates are the $Z_b^+$ particles reported by the Belle collaboration.
Recently, static systems of two quarks and two antiquarks, were studied in the lattice \cite{Alexandrou:2004ak,Okiharu:2004ve}
using an operator with the quantum numbers of the tetraquark.
The results indicate that the static potential of this system could be well described by the generalized flip-flop potential.
This potential, is similar to the ones which were first proposed
\cite{Oka:1984yx,Oka:1985vg} as a device to suppress the non-physical Van der Waals forces,
which arise when potential models based on Casimir Scaling are used, differing by the existence of a third branch of the potential ---
the tetraquark branch.

\section{Tetraquark Wilson Loop}

The Static potential for the tetraquark has been studied in the lattice by \cite{Alexandrou:2004ak} and
\cite{Okiharu:2004ve}.

\begin{figure}
	\label{tetraq_wloop}
	\centering
	\includegraphics[width=0.65\textwidth]{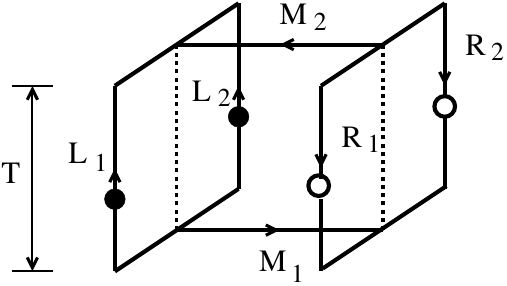}
	\caption{Wilson loop operator used by \cite{Alexandrou:2004ak} and \cite{Okiharu:2004ve}
	to calculate the static potential of a tetraquark system.}
\end{figure}

For that, a Wilson Loop operator was constructed. It is given by

\begin{equation}
	W_{4Q} = \frac{1}{3} \mbox{Tr}[ M_1 R_{12} M_2 L_{12} ] ,
\end{equation}

with

\begin{eqnarray}
	R_{12}^{ii'} &=& \epsilon_{ijk} \epsilon_{i'j'k'} R_1^{jj'} R_2^{kk'} \\ \nonumber
	L_{12}^{ii'} &=& \epsilon_{ijk} \epsilon_{i'j'k'} L_1^{jj'} L_2^{kk'} \, .
\end{eqnarray}
The corresponding paths are visually described on figure \ref{tetraq_wloop}.

This Wilson Loop has the quantum numbers of colour singlet system where
the two quarks form an antitriplet and the two antiquarks form a triplet.

\begin{figure}
	\label{tetraq_string}
	\centering
	\includegraphics[width=0.3\textwidth]{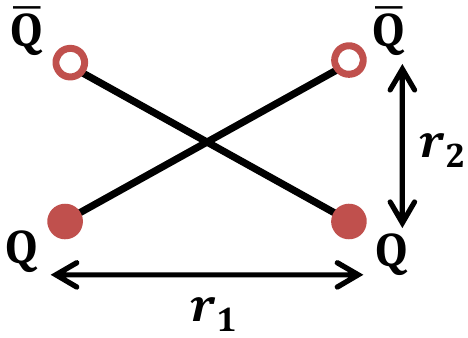}
	\includegraphics[width=0.3\textwidth]{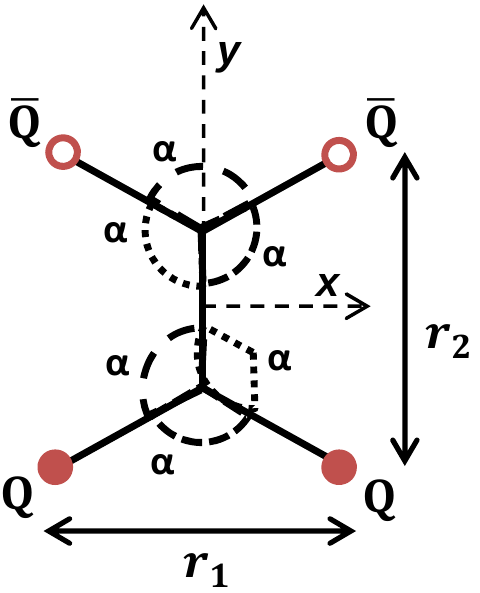}
	\caption{Here we can see the shape of the minimal string which links the four particles
	in the tetraquark. Note that when the diquark and the diantiquark are close the five segment
	structure collapses into a four segment one.}
\end{figure}

This lattice studies, indicate that the static potential of the two-quark
and two-antiquark system is a generalized flip-flop potential:
\begin{equation}
	V_{FF} = \min( V_T , V_{M_1 M_2}, V_{M_3 M_4} )
\end{equation}
where $V_{M_1 M_2}$ and $V_{M_3 M_4}$ are the two possible two-meson
potentials, given by the sum of two independent intra-meson potentials
$V_{M_1 M_2} = V_{M_1} + M_{M_2}$, and $V_T$ is the tetraquark potential
which corresponds to the sector where the four particles are confined,
linked by a single fundamental string. It is given by
\begin{equation}
	V_T = C + \alpha \sum_{i<j} \frac{\lambda_i}{2} \cdot \frac{\lambda_j}{2} + \sigma L_{min} \, ,
\end{equation}
where $L_min$ is the minimal distance linking the four particles, which corresponds to the string configuration displayed on fig. \ref{tetraq_string}.

For the special case where the four particles form as rectangle as in
fig. \ref{tetraq_string}, $L_{min}$ is given by
$L_{min} = \sqrt{3} r_1 + r_2$ for $r_2 > \frac{r_1}{\sqrt{3}}$
\cite{Bicudo:2010mv,Bicudo:2008yr}. If we neglect the Coulomb part of $V_{FF}$, we
could estimate that the tetraquark sector becomes the ground state when
$r_2 \gtrsim \sqrt{3} r_1$.

\section{Tetraquark Fields}

The fields in the tetraquark system can be calculated by using the correlation of plaquettes and Wilson Loops.
The Squared chromoelectric and chromomagnetic fields, are computed by
\begin{eqnarray}
\langle E_{i}^{2}\rangle &=& \langle P_{0i}\rangle-\frac{\langle WP_{0i}\rangle}{\langle W\rangle} \\ \nonumber
\langle B_{i}^{2}\rangle &=& \frac{\langle WP_{jk}\rangle}{\langle W\rangle}-\langle P_{jk}\rangle
\end{eqnarray}
where the index $i$ complments $j$ and $k$. $P_{\mu\nu}$ is
$P_{\mu\nu}=1-\frac{1}{3}\mbox{Tr}[U_{\mu}(\mathbf{s})U_{\nu}(\mathbf{s}+\hat{\nu})U_{\mu}^{\dagger}(\mathbf{s}+\hat{\nu})U_{\nu}^{\dagger}(\mathbf{s})]$.

The Lagrangian and energy densities are given by
\begin{eqnarray}
\langle\mathcal{L}\rangle &=& \frac{1}{2}\langle E^{2}-B^{2}\rangle \\ \nonumber
\langle\mathcal{H}\rangle &=& \frac{1}{2}\langle E^{2}+B^{2}\rangle \, .
\end{eqnarray}

Now, we will present results for the chromo-fields for the tetraquark system for the simple case where the four particles form a
rectangles, as in fig. \ref{tetraq_string}.

This results, and the one of the following chapters were obtained using quenched lattice QCD configurations with dimension $24^3 \times 48$ and
$\beta = 6.2$. They where used in GPUs by using a combination of Cabbibo-Marinari, pseudo heat-bath and over-relaxation algorithms \cite{ptqcd}.
APE smearing \cite{Cardoso:2009kz} and Hypercubic blocking \cite{Hasenfratz:2001hp} were used to improve the signal to noise ratio.

\begin{figure}
	\label{tetraq_fields}
	\centering
	\includegraphics[width=0.22\textwidth]{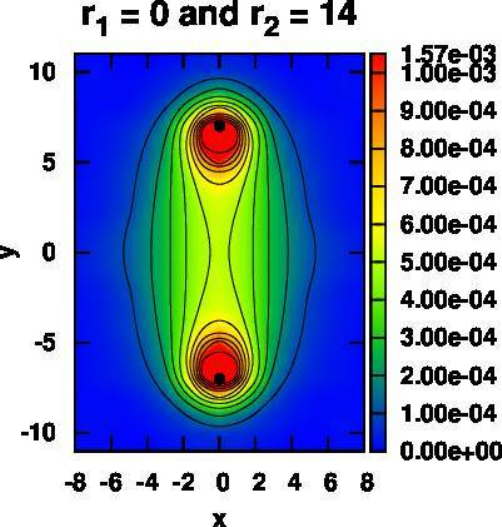}
	\includegraphics[width=0.22\textwidth]{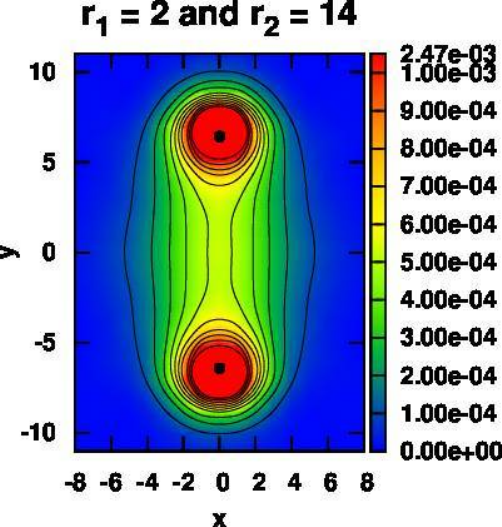}
	\includegraphics[width=0.22\textwidth]{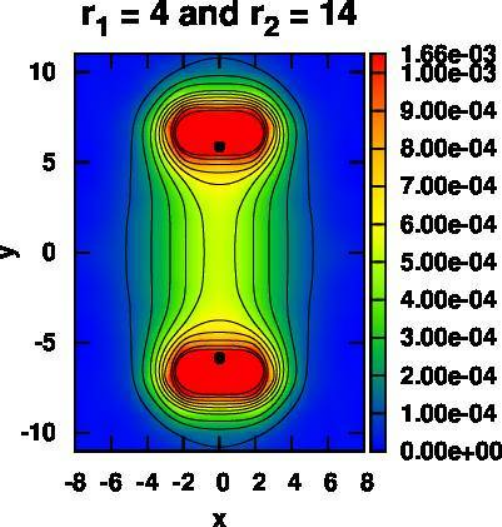}
	\includegraphics[width=0.22\textwidth]{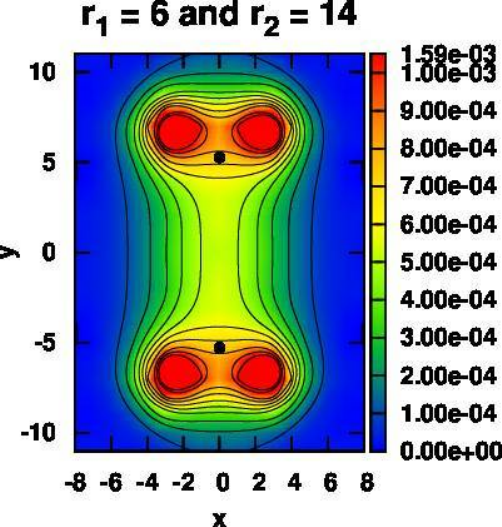}
	\caption{Plot of the lagrangian density for different values of $r_1$ with $r_2 = 14$.}
\end{figure}

The results for the lagrangian density are given on fig. \ref{tetraq_fields}
for fixed $r_2 = 14$.

As we can see the in the leftmost fig. $r_1 = 0$, in which case the system the system is collapsed in to a meson.
Then we can see left to right, with the increase of $r_1$, the confining string acquires a shape similar with the one on
fig. \ref{tetraq_string} for $r_2 > \frac{r_1}{\sqrt{3}}$, in agreement with the result obtained for the static potential.

\begin{figure}
	\centering
	\label{tetraq_fluxtube}
	\includegraphics[width=0.55\textwidth]{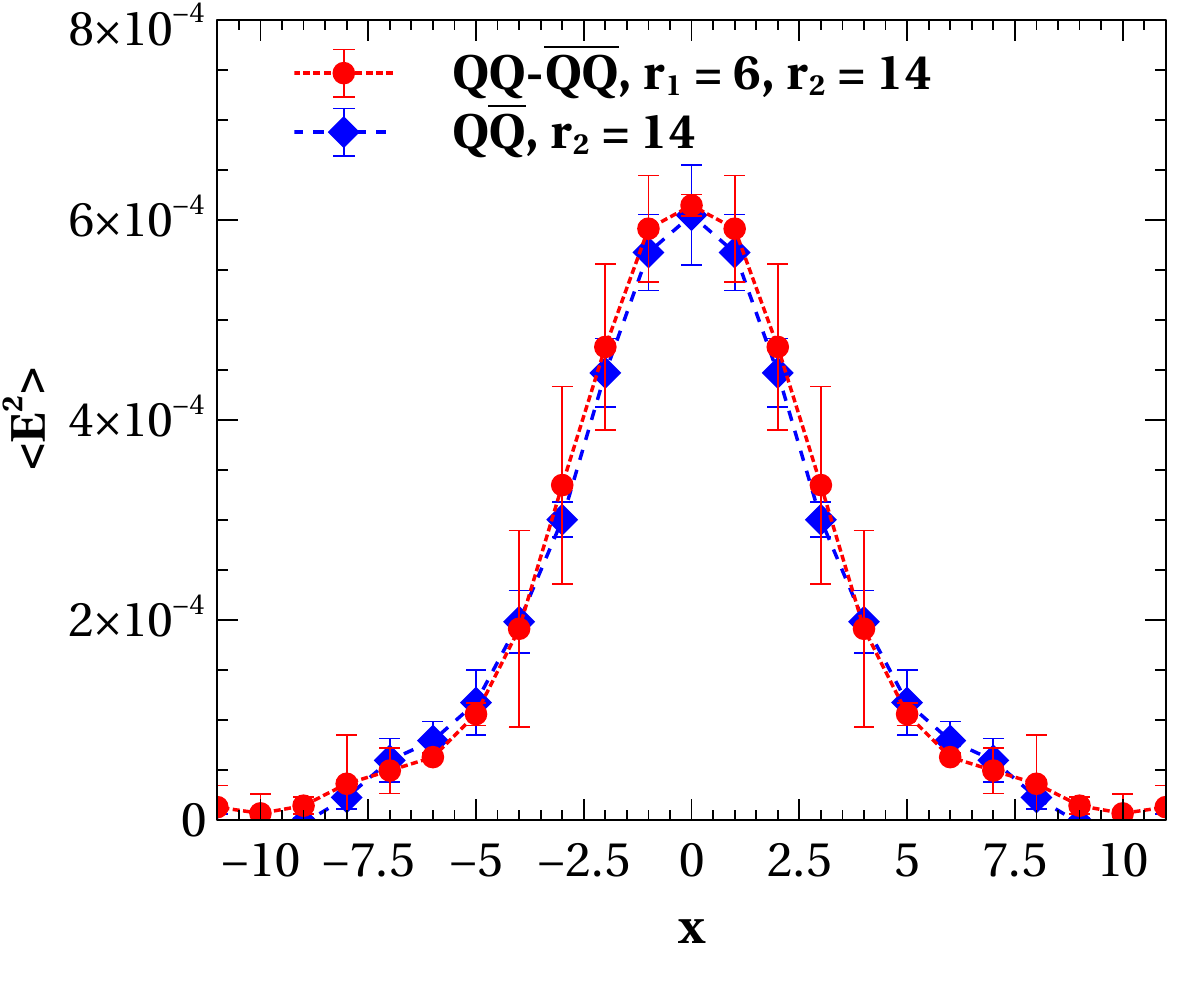}
	\caption{Comparision between the squared chromoelectric field in the quark-antiquark flux tube core (in the meson)
	and the diquark-antidiquark flux tube core (in the tetraquark).}
\end{figure}

In fig. \ref{tetraq_fluxtube} it is compared the squared chromoelectric fields $\langle \mathbf{E}^2 \rangle$ in the
flux tube centre of the meson and in the centre of the diquark-diantiquark flux tube in the tetraquark.
As can be seen both the flux-tubes have a similar behaviour, which confirms that the string which is present
in this system is a fundamental one, again in agreement with the results
for the static potential.

\begin{figure}
	\centering
	\label{tetraq_error}
	\includegraphics[width=0.40\textwidth]{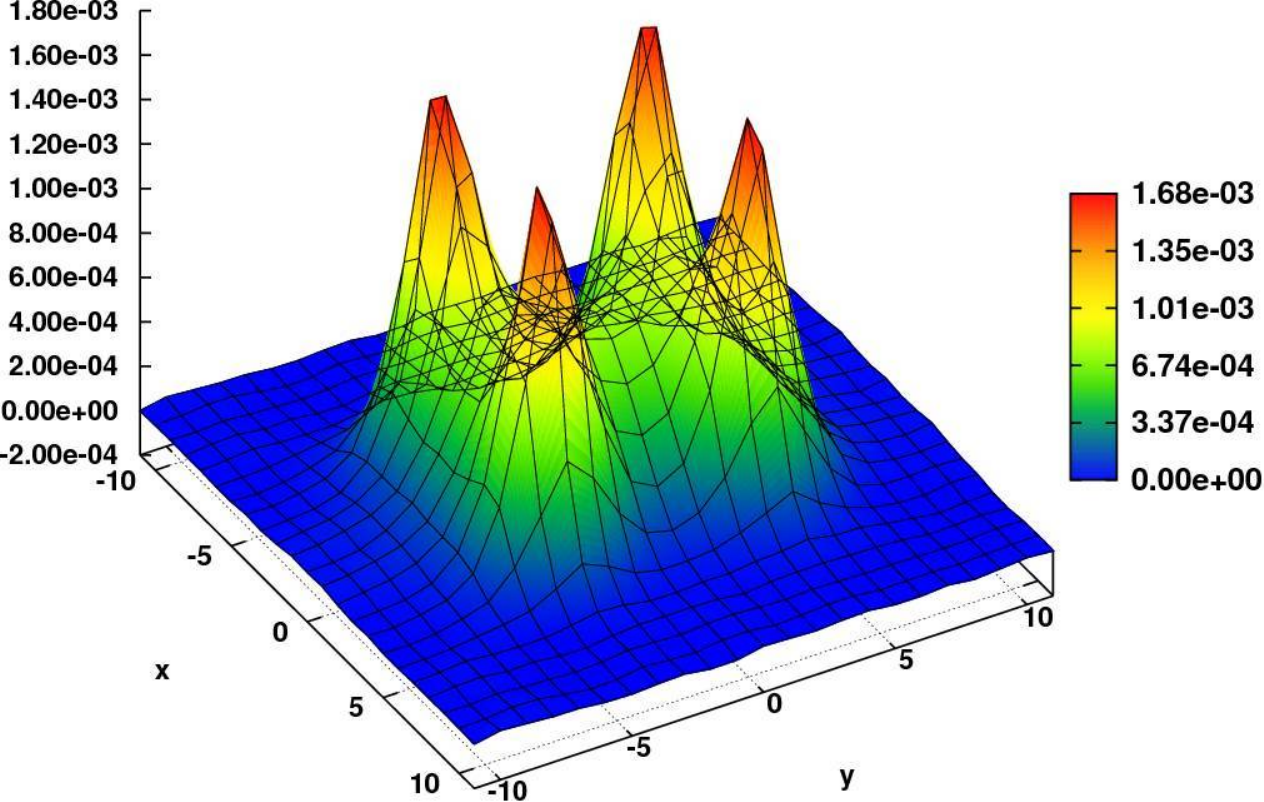}
	\includegraphics[width=0.40\textwidth]{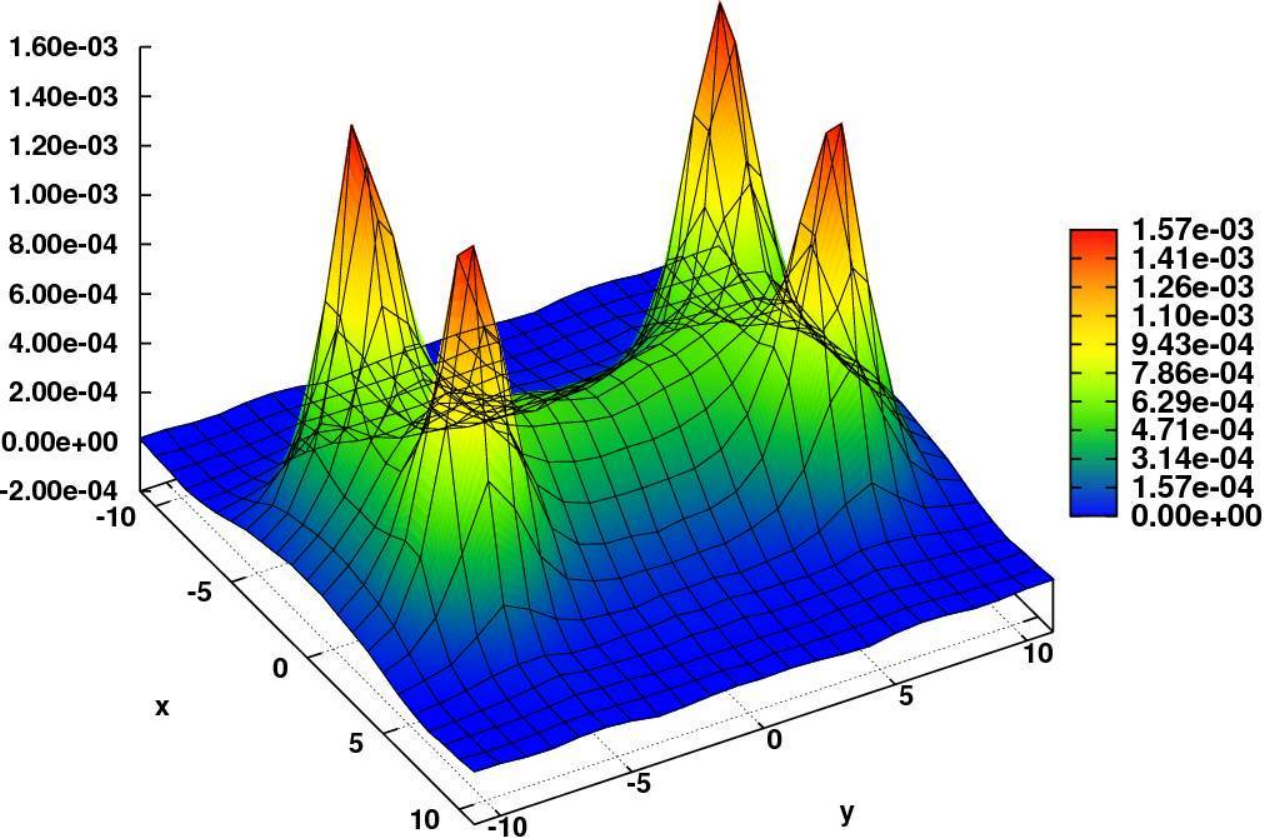}
	\caption{Lagrangian density in the flux tube for $r_1 = r_2 = 8$ and $r_1 = 8, r_2 = 14$.  }
\end{figure}

However, as can be seen in fig. \ref{tetraq_error} this agreement disappears for  $r_2 < \sqrt{3} r_1$ --- the region where the ground
state of the system is a two meson state and not the tetraquark state.
As can be seen there, for $r_1 = 8$ and $r_2 = 14$ (tetraquark region), we obtained the expected ground state.
However for $r_1 = r_2 = 8$ (meson-meson region) we still see a string linking the four particles
as in the tetraquark state, but we should see, to be in agreement with the generalized flip-flop model supported by the
results obtained for the static potential, two strings corresponding
to the two mesons.
We think that this disagreement is due to the low overlap of the used operator with the ground state, and also
to the small temporal extensions used in the Wilson Loop.
To obtain the true ground state of the system in this region and also the first excited state we used a variational method.

\section{Variational Method for the $QQ\bar{Q}\bar{Q}$ system}

Now we will improve this method in order to obtain the true ground state of the system and also to obtain the first
excited state. To achieve this, we note that the Wilson Loop operator could be written as a correlation of a certain
operator at different times $W(t)=\langle \hat{\mathcal{O}}(t) \hat{\mathcal{O}}^\dagger(0) \rangle$. This could be
generalized by considering instead a base of operators $\hat{\mathcal{O}}_i$. So this way, our wilson loop becomes
a matrix $W_{ij} = \langle \hat{\mathcal{O}_i} \hat{\mathcal{O}}^\dagger_j \rangle$. This could be used not only
to improve the ground state overlap but also to obtain more energy levels of the system. These are given by solving the
generalized eigensystem
\begin{equation}
	\langle W_{ij}(t) \rangle c_n^j(t) = w_n \langle W_{ij}(0) \rangle c_n^j(t)
\end{equation}
The fields can then be calculated by
$\langle P \rangle_n =
\frac{\langle W_n P \rangle}{\langle W_n \rangle} - \langle P \rangle$
with $W_n = c_n^i W_{ij} c_n^j$.

\begin{figure}
	\label{geom}
	\centering
	\includegraphics[width=0.35\textwidth]{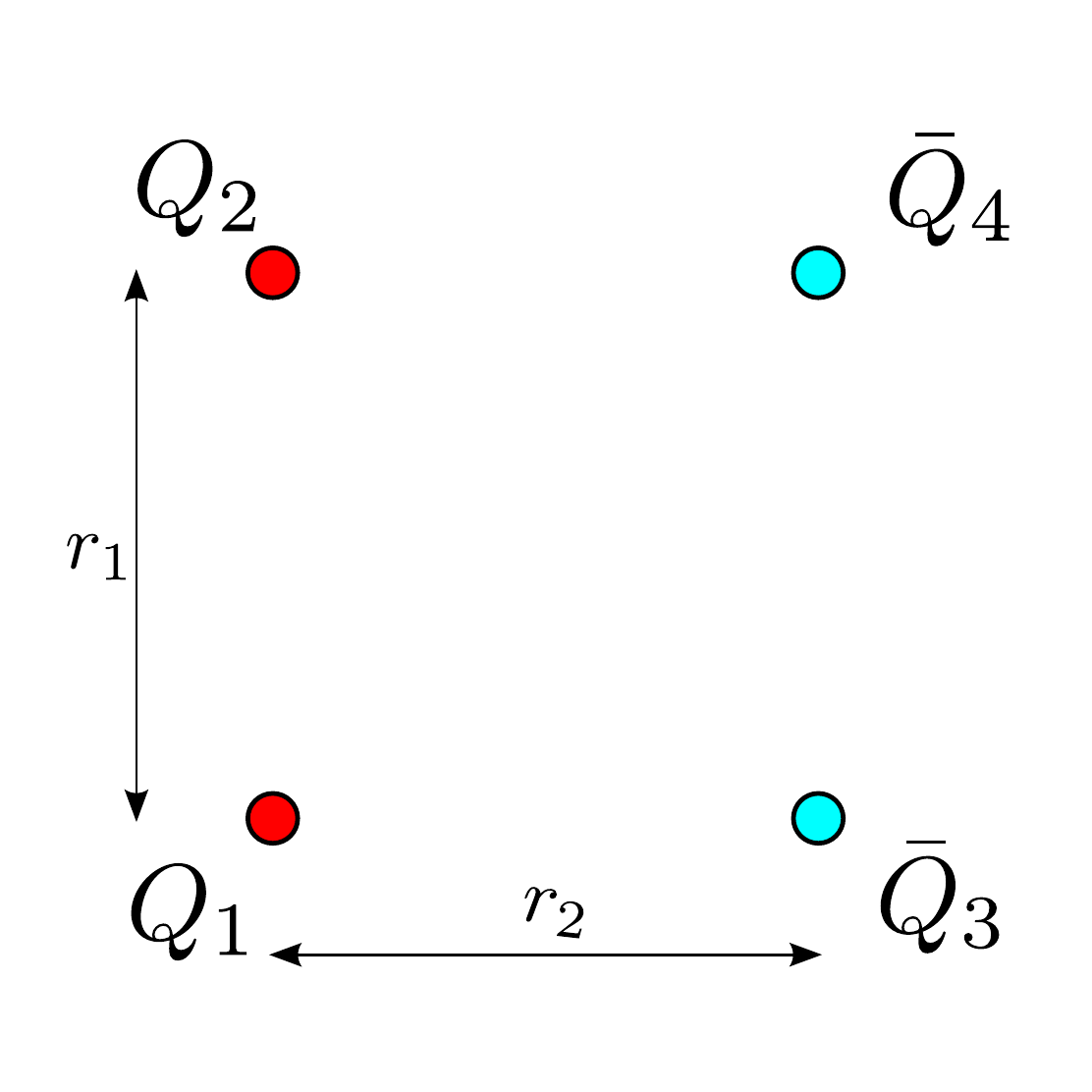}
	\includegraphics[width=0.35\textwidth]{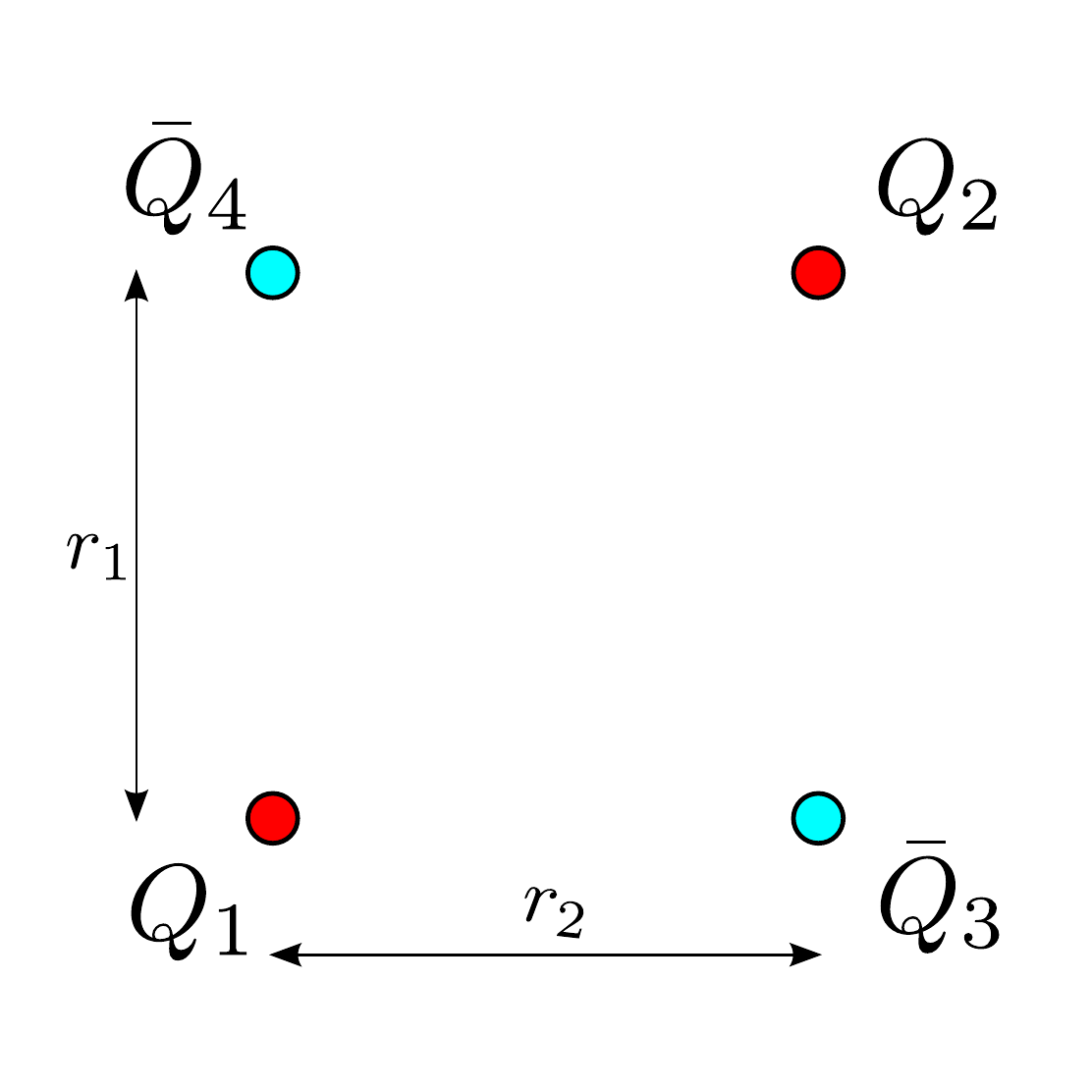}
	\caption{Left: Parallel alignment. Right: Antiparallel alignment.}
\end{figure}

We will consider again the case where the four particles form a rectangles 
and also two different alignments of the $QQ\bar{Q}\bar{Q}$
system. A parallel one, where the two quarks are on the same side of the rectangle and an antiparallel one,
where the quarks are on opposite corners of the rectangle (fig. \ref{geom}).

\begin{figure}
	\label{para_loop}
	\centering
	\includegraphics[width=0.80\textwidth]{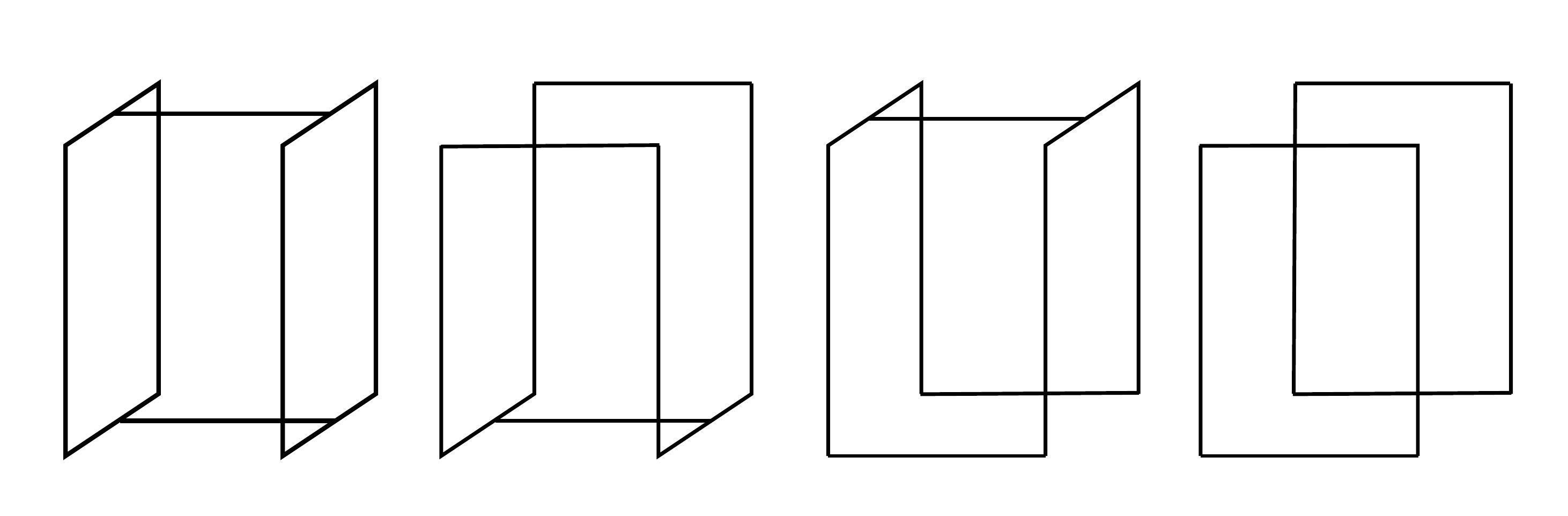}
	\caption{Elements of the Wilson loop matrix used for the parallel alignment.}
\end{figure}

For both cases we will use a base of two operators, similarly to what was made in \cite{Bornyakov:2005kn}.
In the parallel case the operators will be the tetraquark operator $\hat{\mathcal{O}}_{4Q}$, the correlation of which
is the tetraquark Wilson Loop $W_{4Q} = \langle \hat{\mathcal{O}}_{4Q} (t) \hat{\mathcal{O}_{4Q}} (0) \rangle$, and a two meson operator.
This gives a Wilson loop matrix where the diagonal elements are
$W_{4Q}$ and the correlation of two wilson loops, while the off-diagonal
elements correspond to the transition between the two states
(see fig. \ref{para_loop}

\begin{figure}
	\label{antip_loop}
	\centering
	\includegraphics[width=0.80\textwidth]{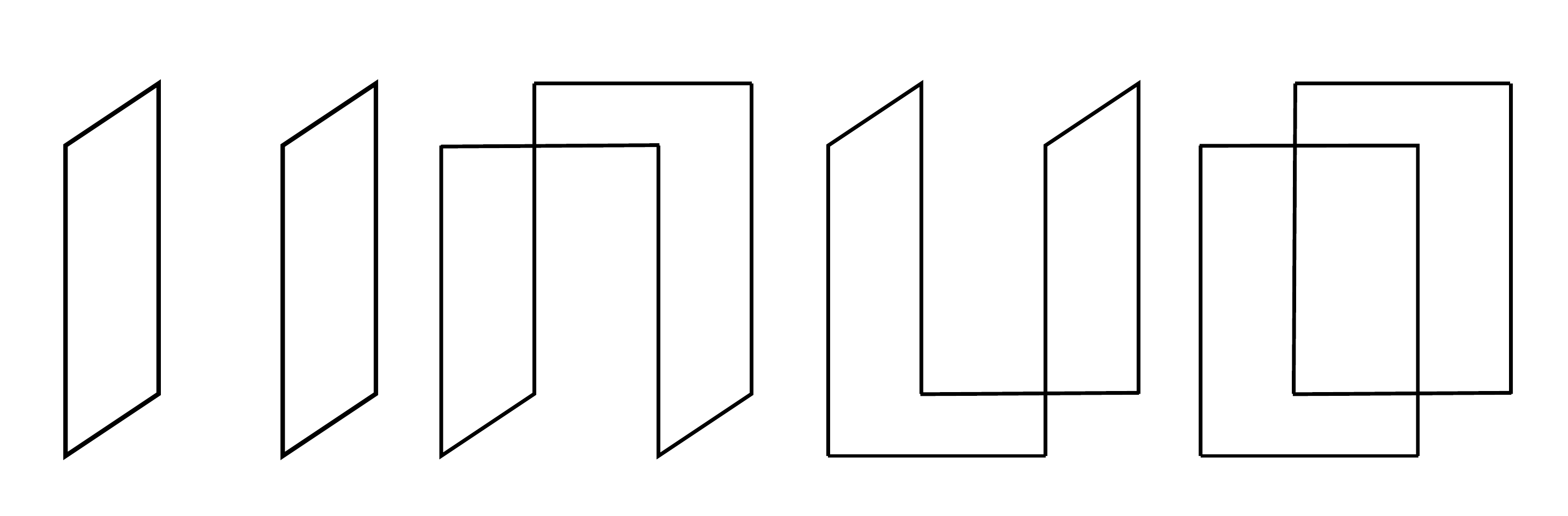}
	\caption{Elements of the Wilson loop matrix used for the antiparallel alignment.}
\end{figure}

For the antiparallel alignment, the operators used were the two meson-meson operators, giving the four matrix elements
given on fig. \ref{antip_loop}.

\section{Results}

Now, we present the results obtained for the parallel geometry.

\begin{figure}
	\label{parallel_components}
	\centering
	\includegraphics[width=0.30\textwidth]{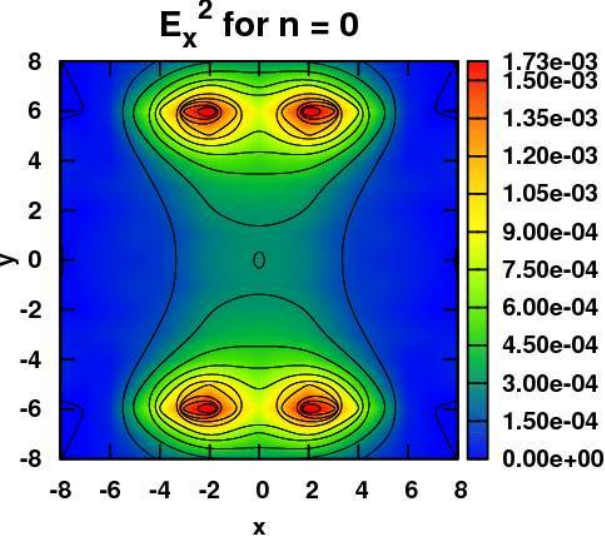}
	\includegraphics[width=0.30\textwidth]{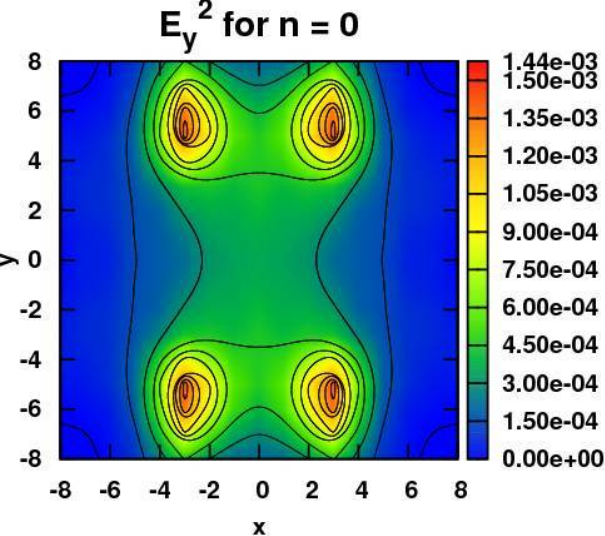}
	\includegraphics[width=0.30\textwidth]{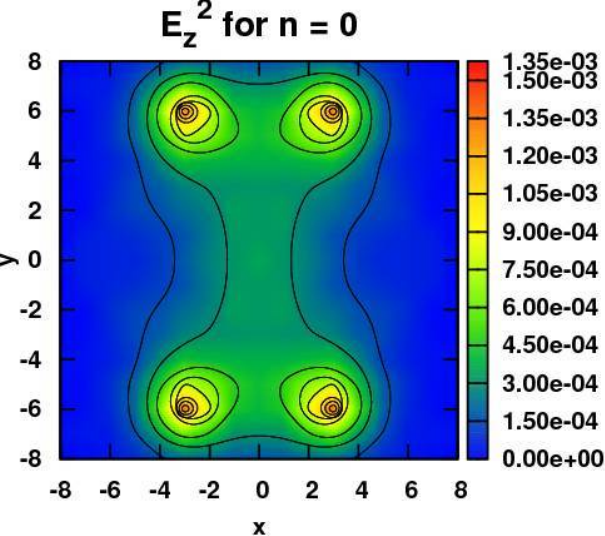}
	\includegraphics[width=0.30\textwidth]{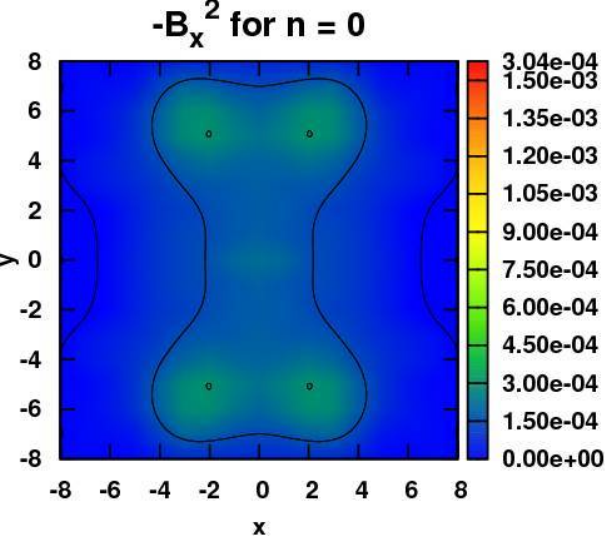}
	\includegraphics[width=0.30\textwidth]{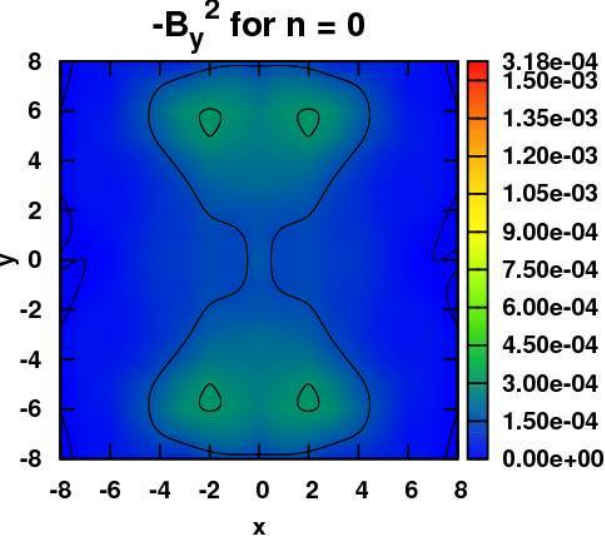}
	\includegraphics[width=0.30\textwidth]{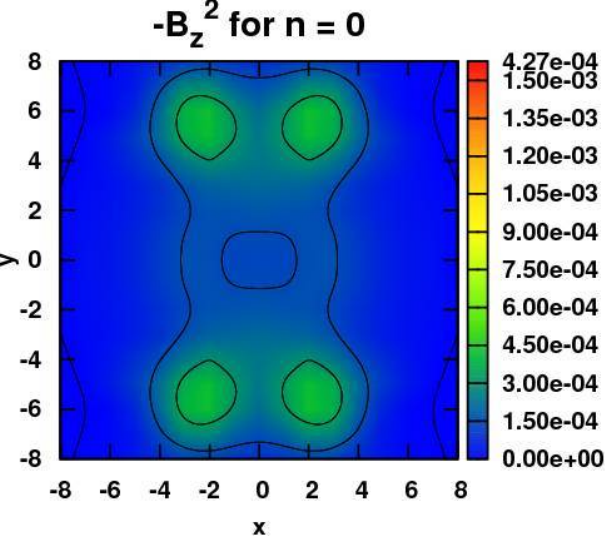}
	\caption{Components of the chromo-electromagnetic field, for the ground state of
	the parallel geometry with $r_1 = 6$ and $r_2 = 12$. }
\end{figure}

\begin{figure}
	\label{antip_components}
	\centering
	\includegraphics[width=0.30\textwidth]{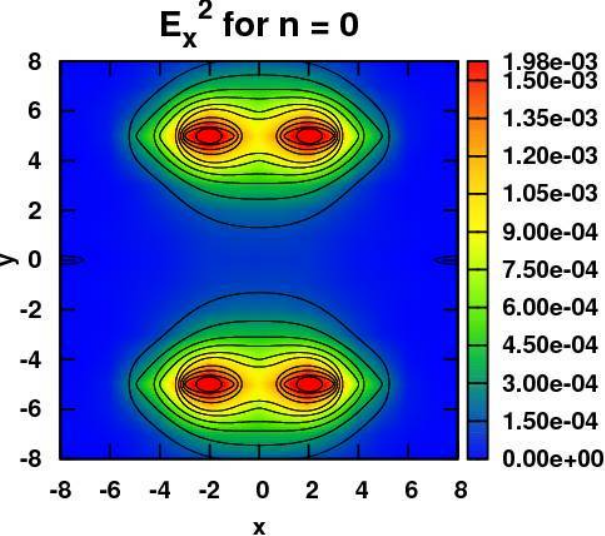}
	\includegraphics[width=0.30\textwidth]{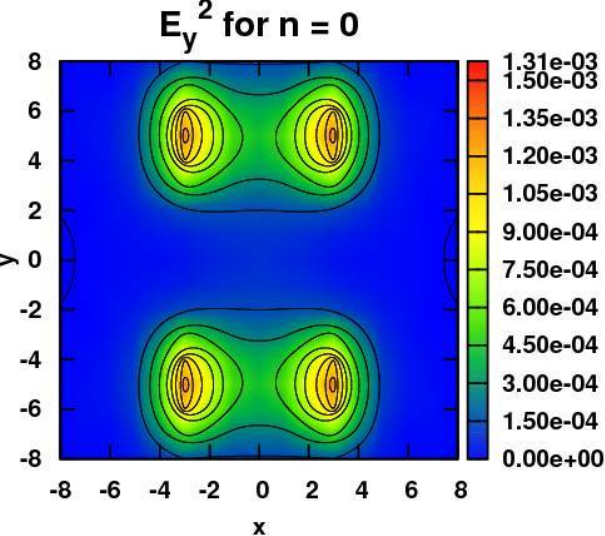}
	\includegraphics[width=0.30\textwidth]{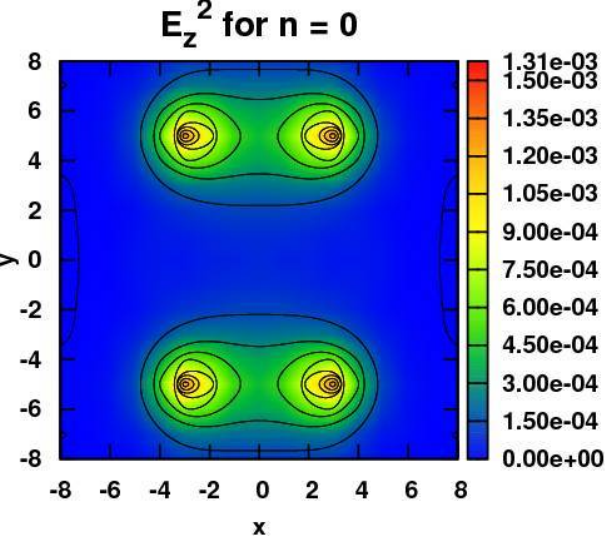}
	\includegraphics[width=0.30\textwidth]{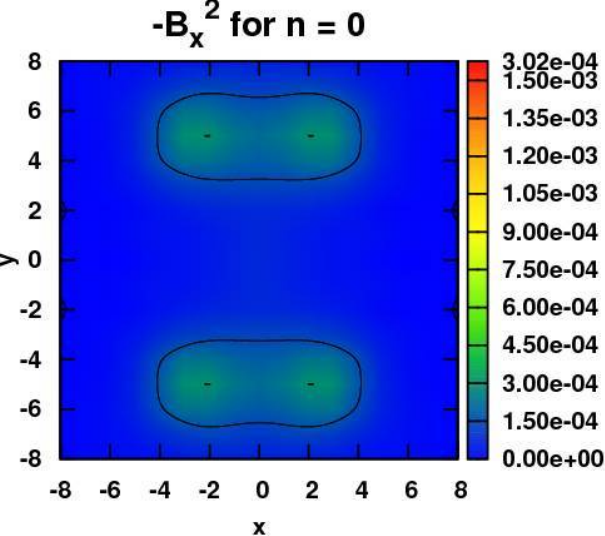}
	\includegraphics[width=0.30\textwidth]{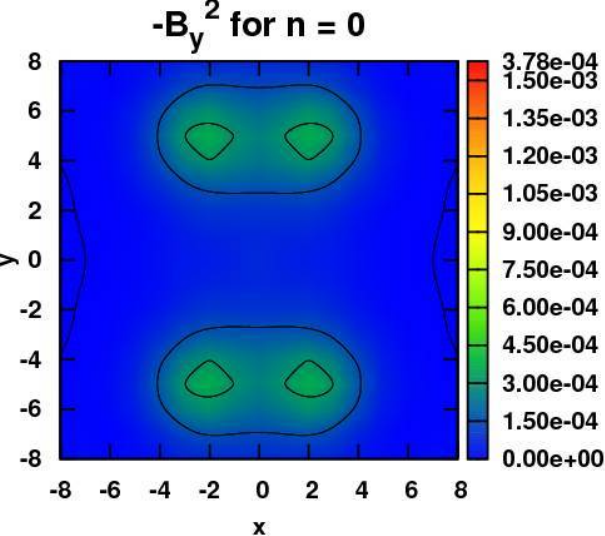}
	\includegraphics[width=0.30\textwidth]{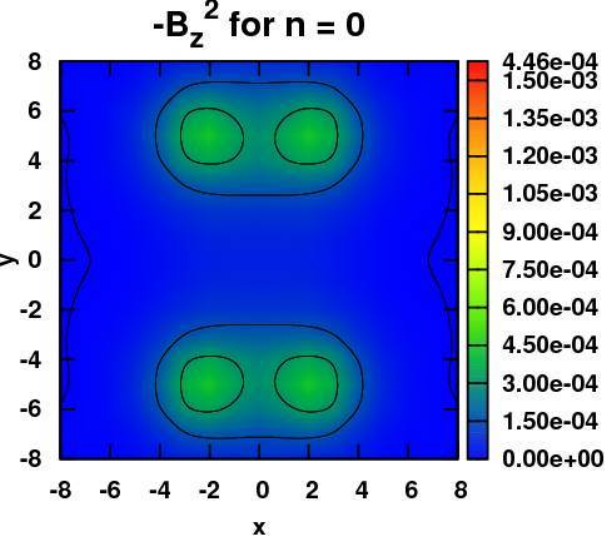}
	\caption{Components of the chromo-electromagnetic field, for the ground state of
	the anti-parallel geometry with $r_1 = 6$ and $r_2 = 10$. }
\end{figure}

In fig. \ref{parallel_components} we can see the components of the chromoelectric and chromomagnetic fields
for the ground state of the parallel alignment, with $r_1 = 6$ and $r_2 = 12$,
while in fig. \ref{antip_components} the componentes are presented for the ground state of the antiparallel alignment
with $r_1 = 6$ and $r_2 = 10$.
The results indicate that what we have is essencialy a tetraquark in the first case and a two meson system in the second.
Besides that we can see features that are common to flux-tubes in general
\cite{Cardoso:2009kz}
such as the dominance of the longitudinal chromoelectric field.

\begin{figure}
	\label{parallel_n0}
	\centering
	\includegraphics[width=0.33\textwidth]{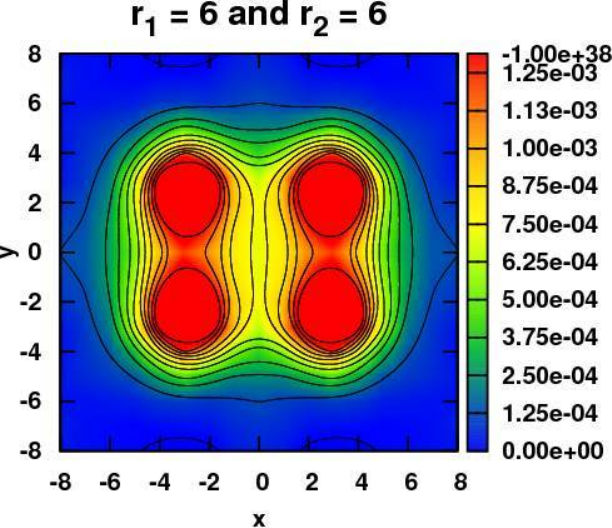}
	\includegraphics[width=0.33\textwidth]{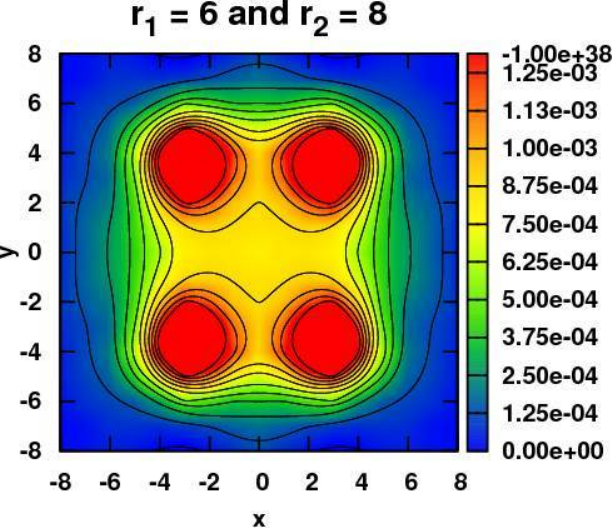}
	\includegraphics[width=0.33\textwidth]{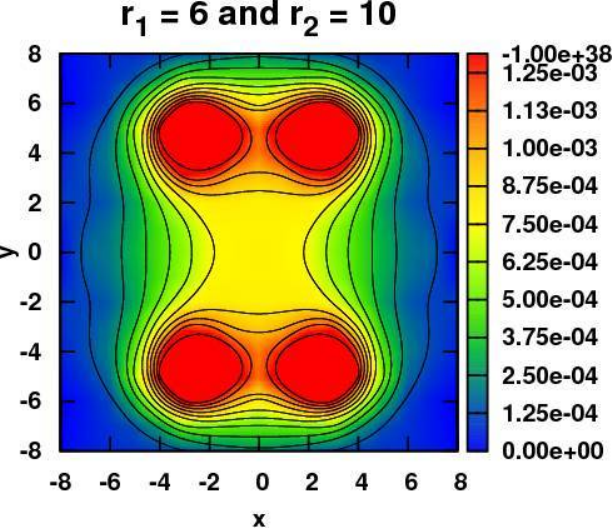}
	\includegraphics[width=0.33\textwidth]{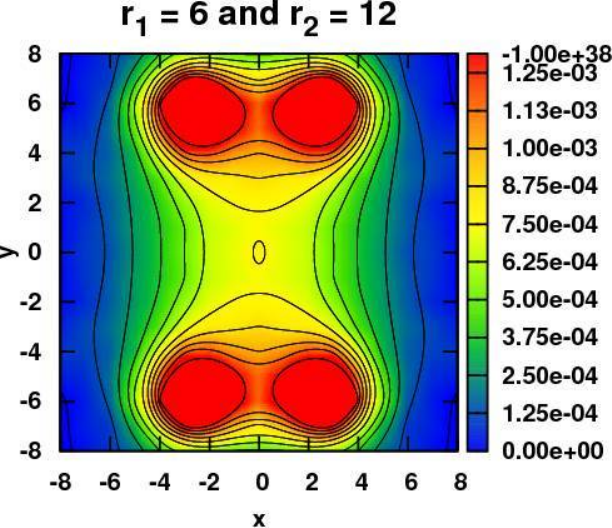}
	\caption{Lagrangian density for different values of $r_1$ and $r_2$ in the ground state of the parallel geometry.}
\end{figure}

\begin{figure}
	\label{parallel_n1}
	\centering
	\includegraphics[width=0.33\textwidth]{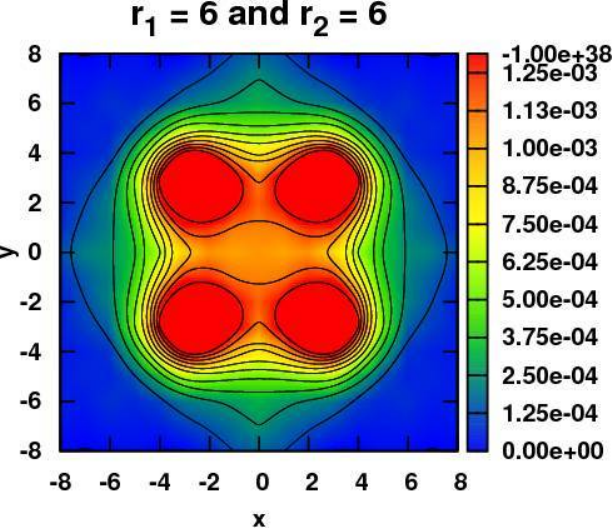}
	\includegraphics[width=0.33\textwidth]{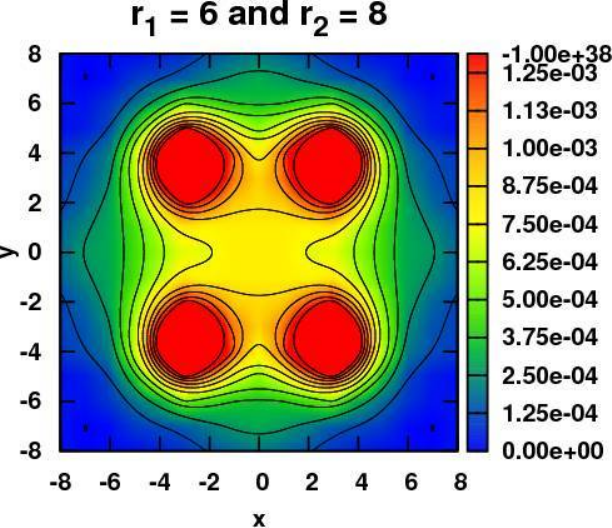}
	\includegraphics[width=0.33\textwidth]{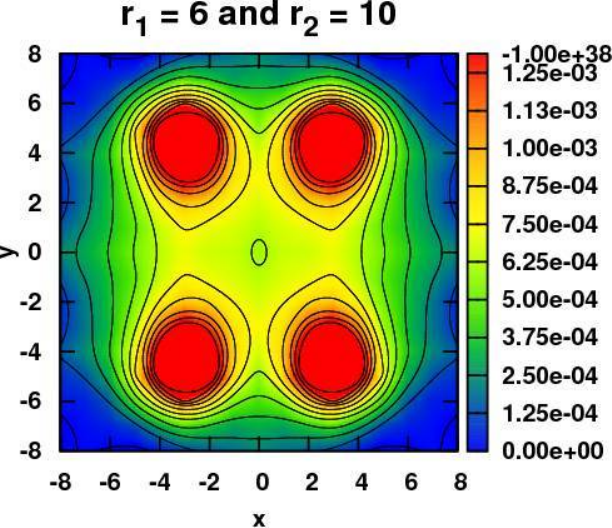}
	\includegraphics[width=0.33\textwidth]{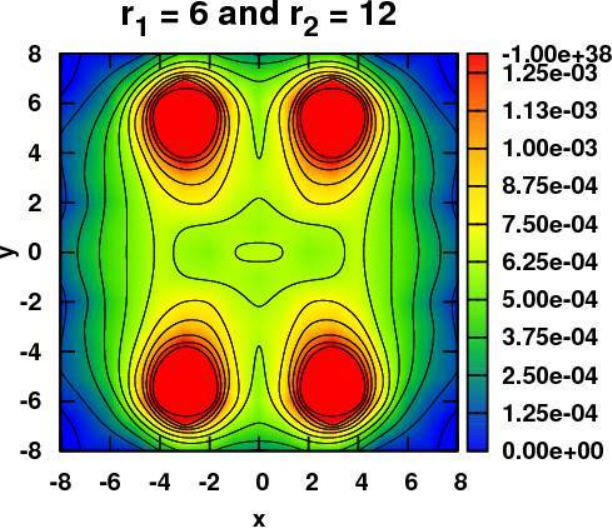}
	\caption{Lagrangian density for different values of $r_1$ and $r_2$ in the first excited state of the parallel geometry.}
\end{figure}

In fig. \ref{parallel_n0} it is shown the evolution of the lagrangian density of the ground state of the parallel geometry,
for fixed $r_1 = 6$.
Here we can see the transition between the two meson state in the top-left picture ($r_2 = 6$)
to the tetraquark state in the bottom right picture $r_2 = 12$.
This is in agreement with the predictions of the generalized flip-flop potential contrary to the results we obtained
using only the tetraquark operator $W_{4Q}$.
The same evolution is presented in \ref{parallel_n1}, but for the first
excited state of the same configurations. This results are not so readily understandable.

\begin{figure}
	\label{parallel_y0}
	\centering
	\includegraphics[width=0.45\textwidth]{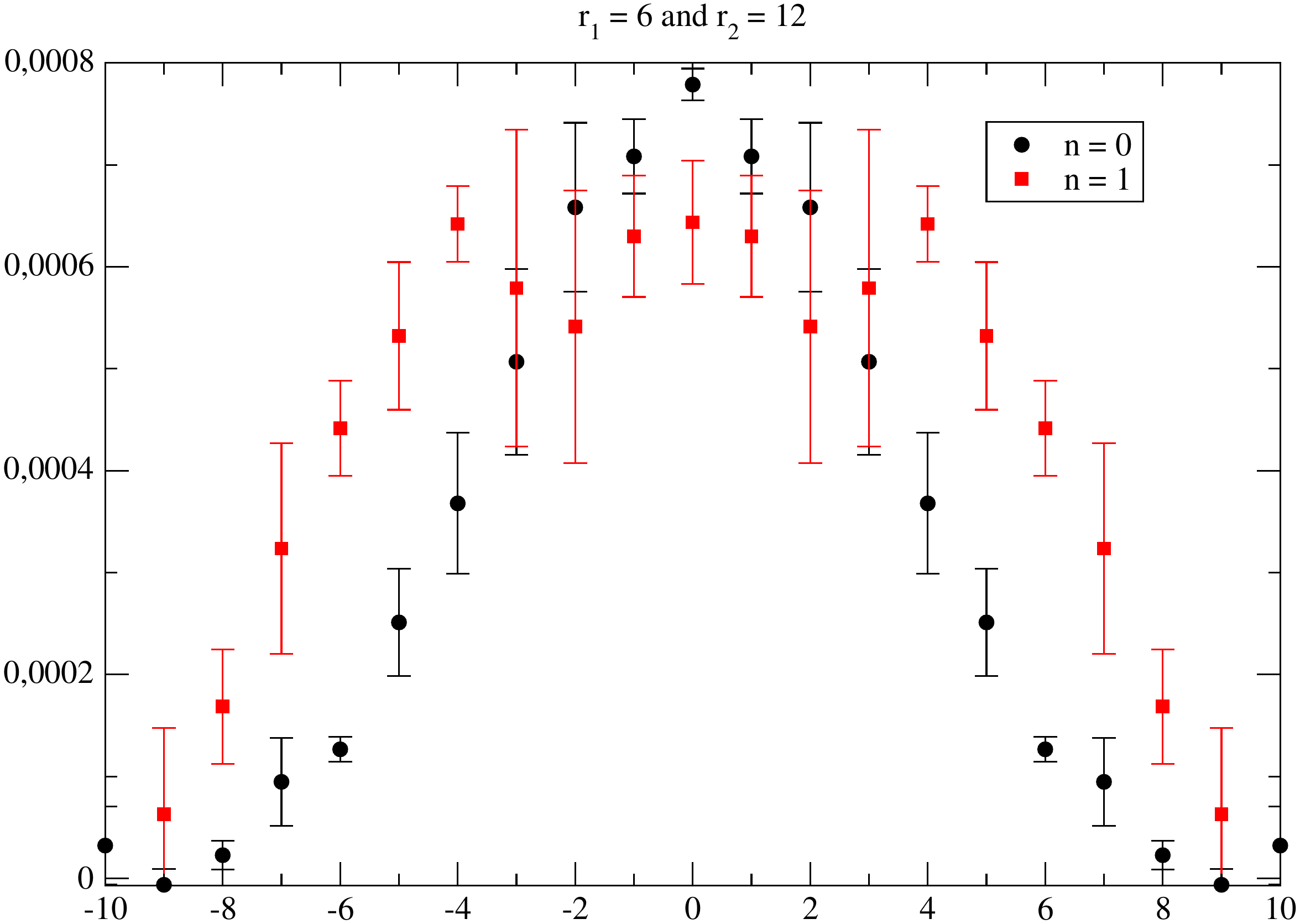}
	\caption{Comparasion of the action density in the center of flux tube in the
	parallel geometry for $r_1 = 6$ and $r_2 = 12$.}
\end{figure}

To better compare the ground and the first excited state we can look at fig. \ref{parallel_y0} in the $y = 0$ axis.

\begin{figure}
	\label{antip_n0}
	\centering
	\includegraphics[width=0.30\textwidth]{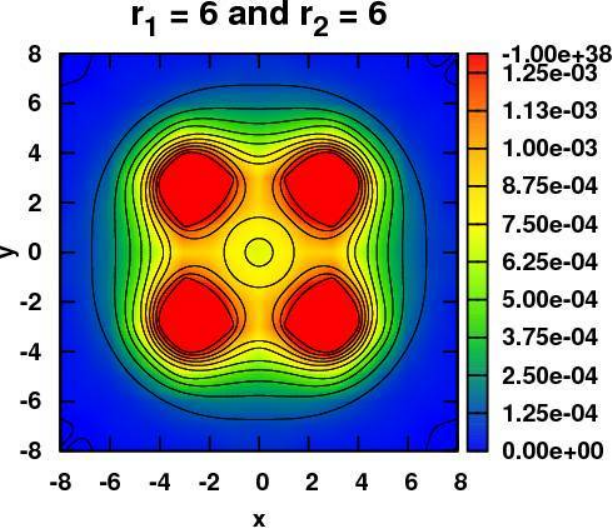}
	\includegraphics[width=0.30\textwidth]{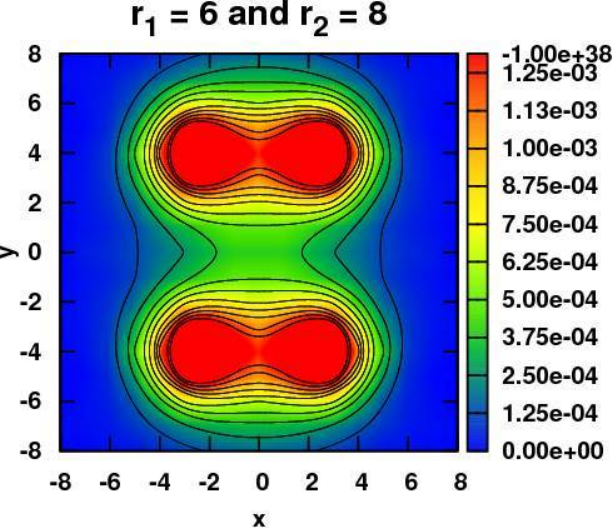}
	\includegraphics[width=0.30\textwidth]{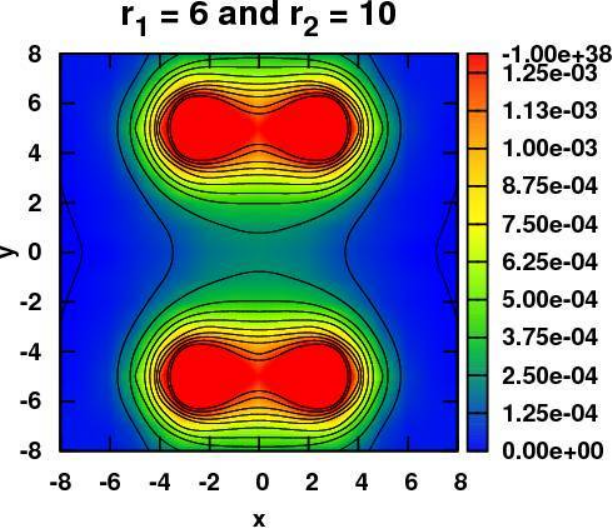}
	\includegraphics[width=0.30\textwidth]{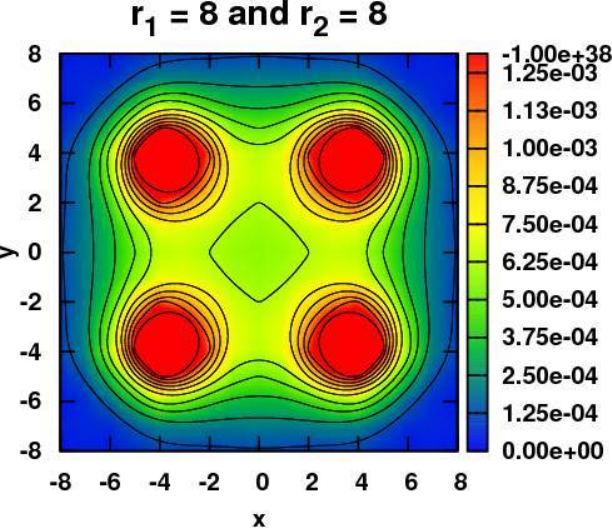}
	\includegraphics[width=0.30\textwidth]{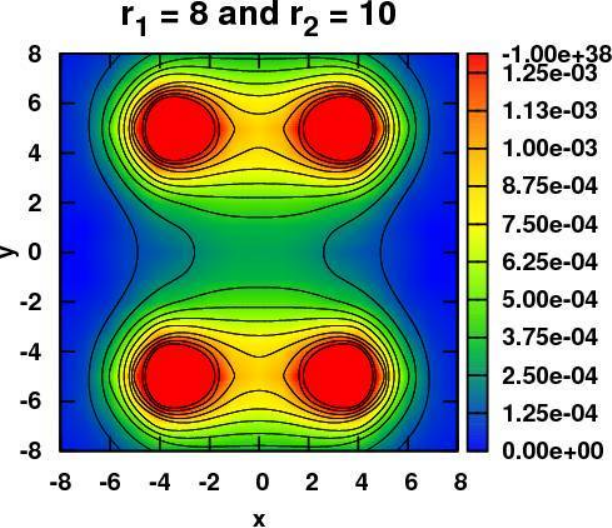}
	\includegraphics[width=0.30\textwidth]{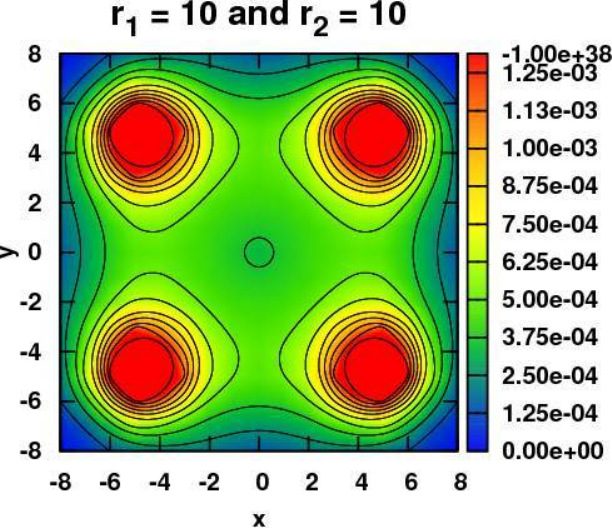}
	\caption{Lagrangian density for different values of $r_1$ and $r_2$ in the ground state of the anti-parallel geometry.}
\end{figure}

In fig. \ref{antip_n0} we see the lagrangian density for the parallel
alignment with varying $r_1$ and $r_2$. For $r_1 = 6$ and $r_2 = 10$
we can see the two mesons almost completely separatedly, while for $r_1 = r_2$ we have a state when all the four particles are linked.
We only show pictures for $r_2 \geq r_1$. The case $r_2 < r_1$ can be seen by rotating the pictures for $r_2 > r_1$.
This behaviour is again expected, as we obtain the separation in two mesons of the system when it correspond to the minimal string energy.
The behaviour for $r_1 = r_2$ is also expected, because neither of the two meson systems is preferred over the other.

\begin{figure}
	\label{antip_n1}
	\centering
	\includegraphics[width=0.30\textwidth]{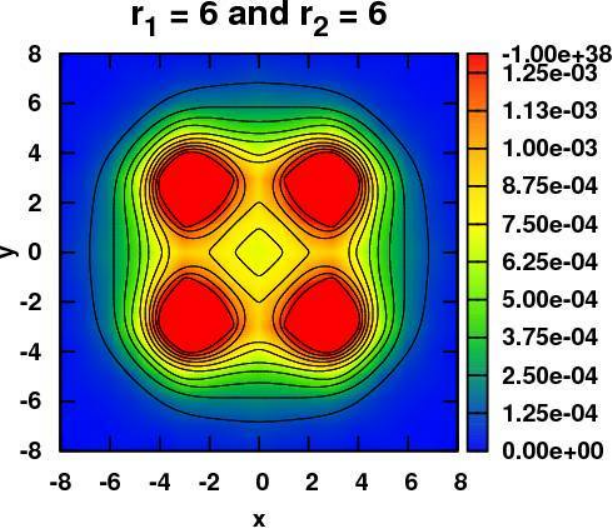}
	\includegraphics[width=0.30\textwidth]{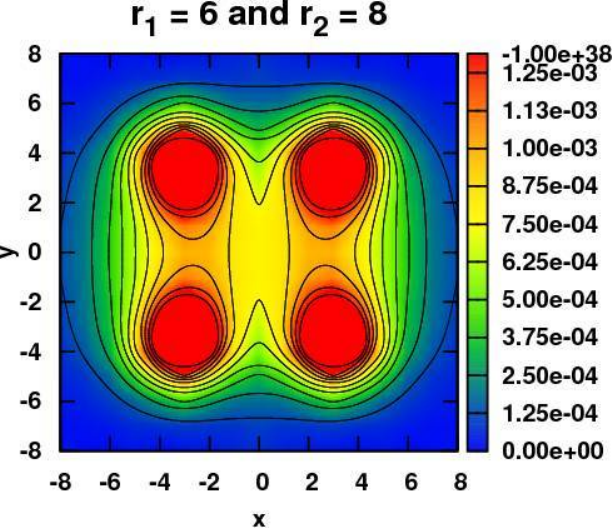}
	\includegraphics[width=0.30\textwidth]{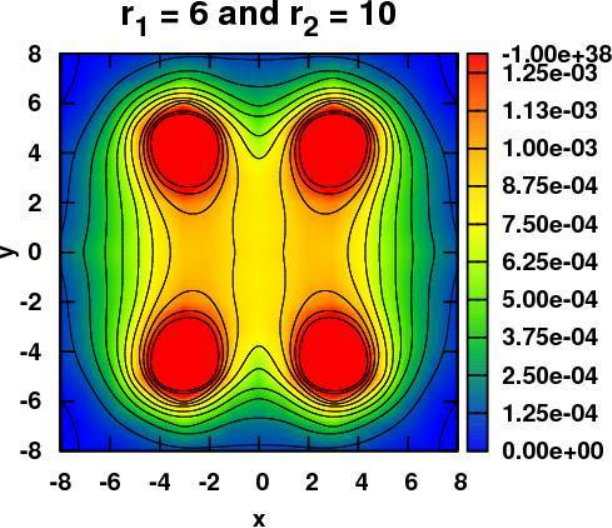}
	\includegraphics[width=0.30\textwidth]{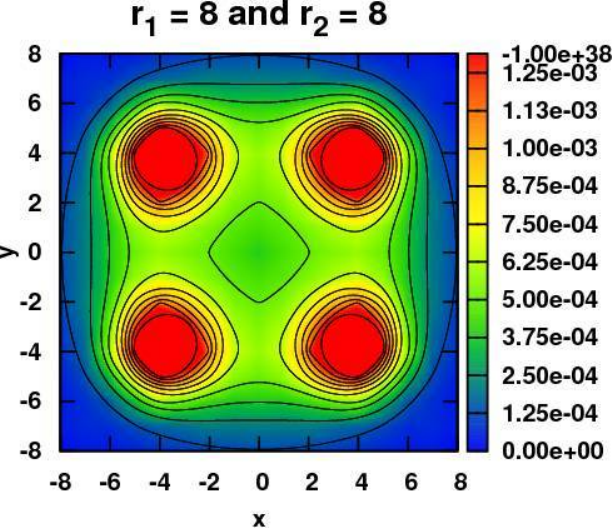}
	\includegraphics[width=0.30\textwidth]{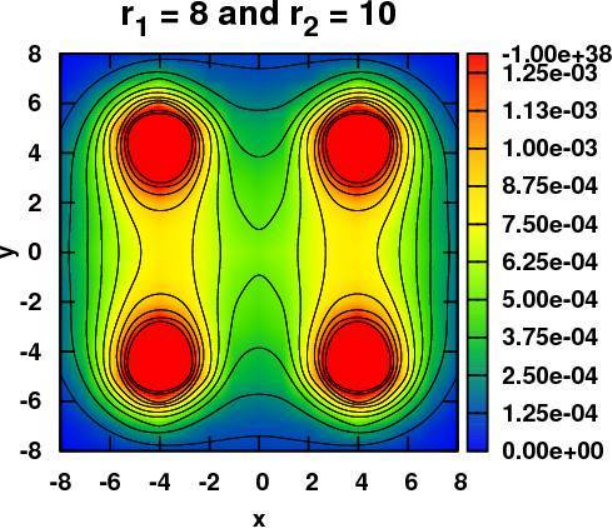}
	\includegraphics[width=0.30\textwidth]{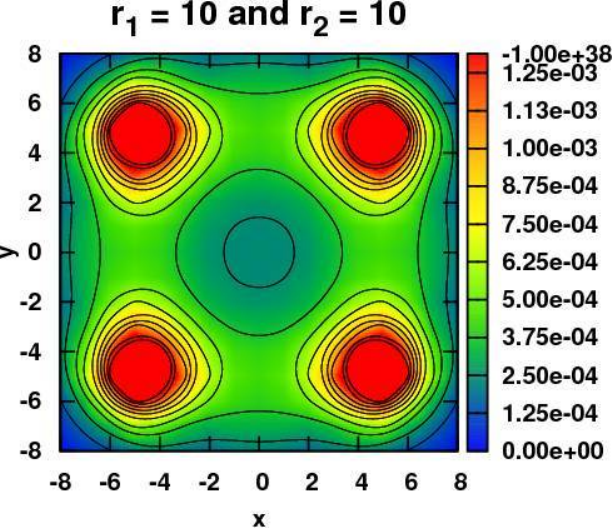}
	\caption{Lagrangian density for different values of $r_1$ and $r_2$ in the first excited state of the anti-parallel geometry.}
\end{figure}

The results for the first excited state are presented in fig.
\ref{antip_n1}.
As we can see there, the first excited state is different from a two meson state, with all the four particles being linked
the flux-tube.

\begin{figure}
	\label{antip_y0}
	\centering
	\includegraphics[width=0.45\textwidth]{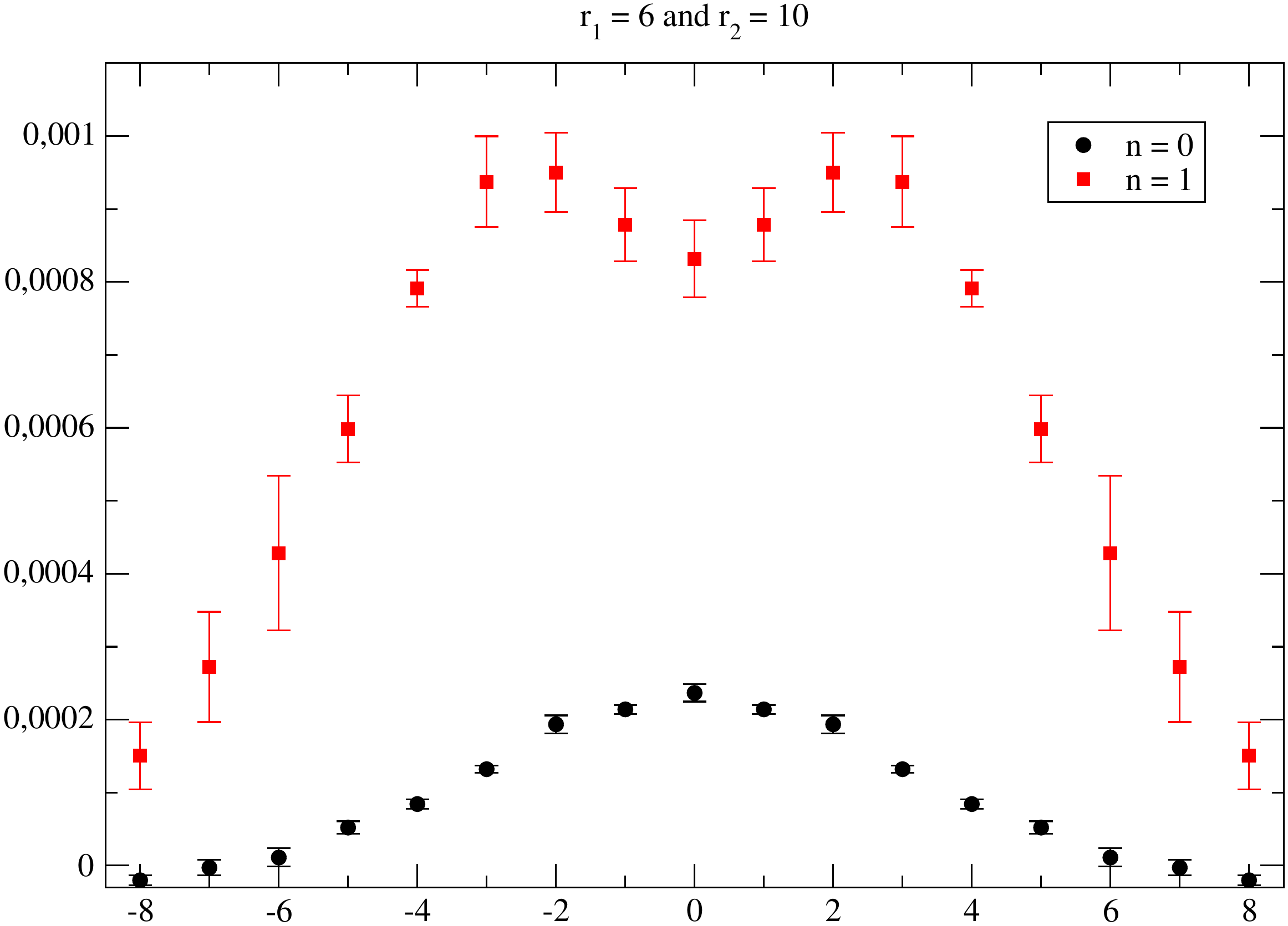}
	\includegraphics[width=0.45\textwidth]{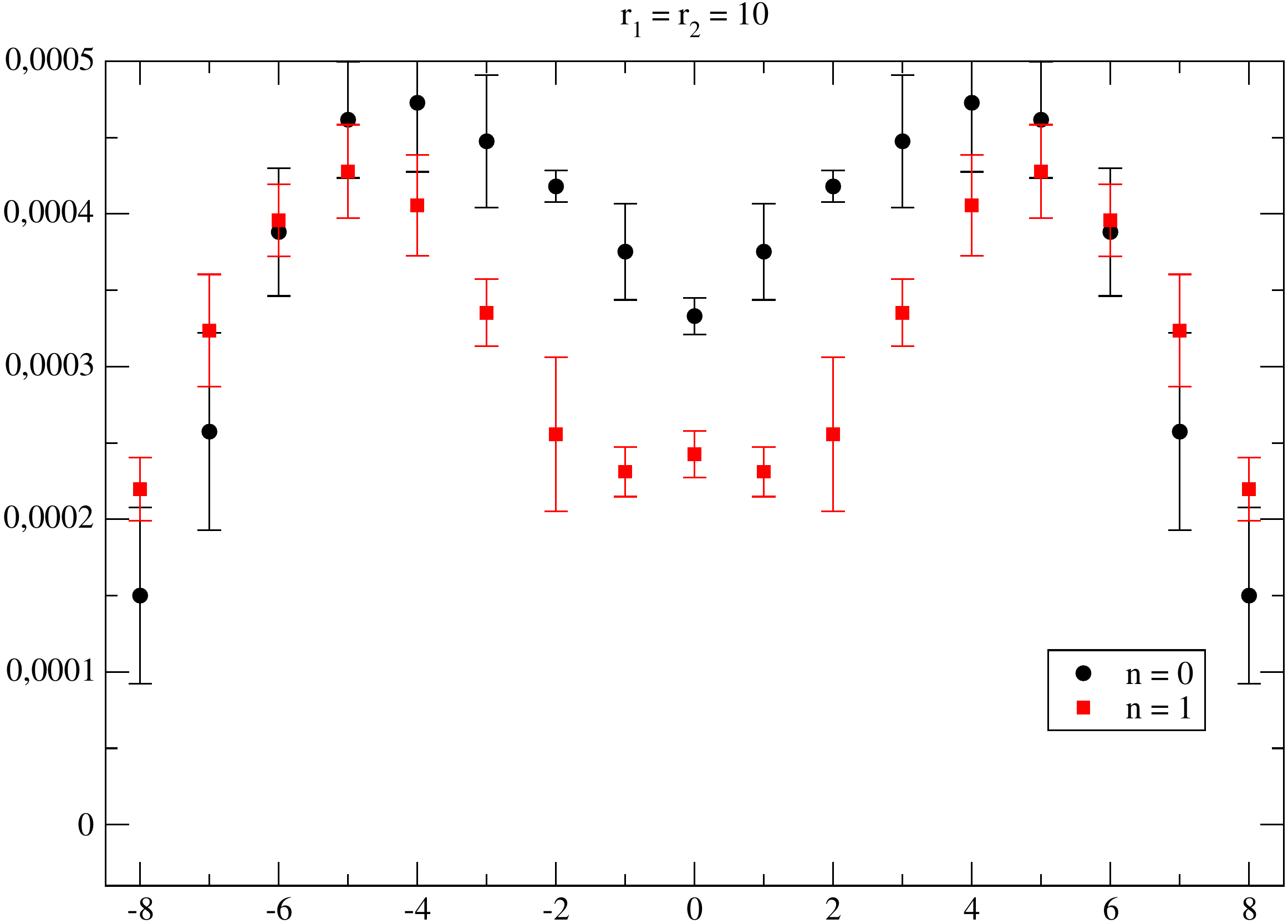}
	\caption{Comparasion of the action density in the center of flux tube in the
	anti-parallel geometry for two geometries.}
\end{figure}

Finally we compare the ground state and the first excited state in this geometry in fig. \ref{antip_y0} for $y = 0$,
in the left, for $r_1 = 6$ and $r_2 = 10$, in the right for $r_1 = r_2 = 10$.
As can be seen, for the first case, the ground state flux-tube in this region is very small,
because of the formation of two mesons.

\section{Discussion}

Our results for the fields of the two quarks and two antiquarks system agree with the results obtained for the static potential, therefore
supporting a string picture of confinement with flux-tube recombination.
Besides this, we were also able to calculate the first excited state of this system. The low overlap between the tetraquark operator and the
two meson ground state may indicate small width tetraquark resonances.
We are also studying the fields of the pentaquarks and the static potentials of this system.

This work was partly funded by the FCT contracts, PTDC/FIS/100968/2008, CERN/FP/109327/2009
and CERN/FP/116383/2010. Nuno Cardoso is also supported by FCT under the contract SFRH/BD/44416/2008 and 
Marco Cardoso is supported under contract SFRH/BPD/73140/2010 by the same institution.

\bibliographystyle{ieeetr}
\bibliography{bib}

%\begin{thebibliography}{99}
%\bibitem{...} 
%\end{thebibliography}

\end{document}